\documentclass[12pt]{article}
\usepackage{amsfonts}
\usepackage{graphicx}
\usepackage{amsmath}
\usepackage{amssymb}
\usepackage{caption2}
\begin{document}
\bibliographystyle{prsty}
\begin{center}
{\large {\bf \sc{ Analysis of the nonet scalar  mesons
as tetraquark states with new  QCD sum rules }}} \\[2mm]
Zhi-Gang Wang  \footnote{E-mail:wangzgyiti@yahoo.com.cn. } \\
  Department of Physics, North China Electric Power University, Baoding 071003, P. R.
  China
\end{center}

\begin{abstract}
In this article, we take the scalar diquarks as point particles and
describe them as  basic quantum fields, then introduce the $SU(3)$
color gauge interaction and new vacuum condensates to study the
nonet scalar  mesons as tetraquark states with the QCD sum rules.
Comparing with the conventional quark currents, the diquark currents
have the outstanding advantage to
 satisfy the two criteria of the QCD sum rules more easily.
\end{abstract}

PACS numbers:  12.38.Lg; 13.25.Jx; 14.40.Cs

{\bf{Key Words:}}  Nonet scalar mesons,  QCD sum rules
\section{Introduction}

The light flavor  scalar mesons present a remarkable exception for
the naive  quark model, and the structures of those mesons have not
been unambiguously determined yet. The  numerous candidates with
$J^{PC}=0^{++}$ below $2\,\rm{GeV}$ cannot be accommodated in one
$q\bar{q}$ nonet, some are supposed to be glueballs, molecular
states and tetraquark states (or their special superpositions)
\cite{ReviewGodfray,Close2002,ReviewScalar}. The $a_0(980)$ and
$f_0(980)$ are good candidates for the $K{\bar K}$ molecular states
\cite{IsgurKK}, however, their cousins $\sigma(600)$ and
$\kappa(800)$ lie considerably higher than the corresponding
thresholds, it is difficult to identify them as the $\pi\pi$ and
$\pi K$  molecular states, respectively.  There may be different
dynamics which dominate the $0^{++}$ mesons  below and above
$1\,\rm{GeV}$ respectively,  and result in two scalar nonets below
$1.7\,\rm{ GeV}$ \cite{ReviewGodfray,Close2002,ReviewScalar}. The
strong attractions between the diquark states $(qq)_{\overline{3}}$
and $(\bar{q}\bar{q})_{3}$  in relative
 $S$-wave may result in a nonet tetraquark states  manifest below
$1\,\rm{GeV}$, while the conventional $^3P_0$  $q\bar{q}$ nonet have
masses about $(1.2-1.6) \,\rm{GeV}$, and the well established
$^3P_1$ and $^3P_2$   $q\bar{q}$ nonets with $J^{PC}=1^{++}$ and
$2^{++}$ respectively lie in the same region. Furthermore, there are
enough candidates for the $^3P_0$   $q\bar{q}$ nonet mesons,
$a_0(1450)$, $f_0(1370)$, $K^*(1430)$, $f_0(1500)$ and $f_0(1710)$
\cite{PDG}.

  In the tetraquark
scenario, the structures of the nonet scalar mesons in the ideal
mixing limit can be symbolically written  as \cite{Jaffe1977}
\begin{eqnarray}
\sigma(600)=ud\bar{u}\bar{d},\;&&f_0(980)={us\bar{u}\bar{s}+
ds\bar{d}\bar{s}\over\sqrt{2}}, \nonumber\\
a_0^-(980)=ds\bar{u}\bar{s},\;&&a_0^0(980)={us\bar{u}\bar{s}
-ds\bar{d}\bar{s}\over\sqrt{2}},\;a_0^+(980)=us\bar{d}\bar{s},
\nonumber\\
\kappa^+(800)=ud\bar{d}\bar{s},\;&&\kappa^0(800)=ud\bar{u}\bar{s},\;\;\;\;\;\;
\bar{\kappa}^0(800)=us\bar{u}\bar{d},
\;\;\;\;\;\;\kappa^-(800)=ds\bar{u}\bar{d} \, .  \nonumber \\
\end{eqnarray}
 The four light
isospin-$\frac{1}{2}$ $K\pi$ resonances near $800 \,\rm{MeV}$, known
as the $\kappa(800)$  mesons,  have not been firmly established yet,
there are still controversy   about their existence due to the large
width and nearby $K\pi$ threshold \cite{PDG}.

In general, we may expect  constructing   the tetraquark currents
and studying  the   nonet scalar  mesons below $1\,\rm{GeV}$ as the
tetraquark states with the  QCD sum rules
\cite{Shifman79,Reinders85}. For the conventional  mesons and
baryons,    the "single-pole + continuum states" model works well in
representing the phenomenological spectral densities, the continuum
states are usually approximated by the contributions from the
asymptotic quarks and gluons, the Borel windows  are rather large
and reliable QCD sum rules can be obtained. However, for the light
flavor multiquark states, we cannot obtain a Borel window to satisfy
the two criteria (pole dominance and convergence of the operator
product expansion) of the QCD sum rules \cite{Narison04}. In
Ref.\cite{Brito2005}, T. V. Brito  et al take the quarks as the
basic quantum fields, and  study the scalar mesons $\sigma(600)$,
$\kappa(800)$, $f_0(980)$ and $a_0(980)$ as the  diquak-antidiquark
states with the QCD sum rules, and cannot obtain  Borel windows to
satisfy the two criteria, and resort to a compromise between the two
criteria. For the heavy tetraquark states and molecular states, the
two criteria can be satisfied, but the Borel windows are rather
small \cite{HTetraquark}.

We can take the colored diquarks as point particles and describe
them as the basic  scalar, pseudoscalar, vector, axial-vector and
tensor fields respectively to overcome the embarrassment
\cite{Wang1003}. In this article,  we construct the color singlet
tetraquark currents with the scalar diquark fields, parameterize the
nonperturbative effects with new vacuum condensates  besides the
gluon condensate, and  perform the standard procedure of the QCD sum
rules to study the  nonet scalar mesons below $1\,\rm{GeV}$. The QCD
sum rules are "new" because the interpolating currents are
constructed  from  the basic diquark fields instead of the quark and
gluon fields.

Whether or not the colored diquarks can be taken as basic
constituents is of great importance, because it provides a new
spectroscopy for the mesons and baryons \cite{Jaffe1977,DiquarkM}.

The article is arranged as follows:  we derive the new QCD sum rules
for  the  nonet scalar  mesons  in Sect.2; in Sect.3, we present the
 numerical results and discussions; and Sect.4 is reserved for our
conclusions.

\section{ The nonet scalar mesons  with  QCD Sum Rules}

In the following, we write down the interpolating currents  for the
nonet scalar mesons below $1\,\rm{GeV}$,
\begin{eqnarray}
J_{f_0}(x)&=&\frac{D^a(x)\bar{D}^a(x)+U^a(x)\bar{U}^a(x)}{\sqrt{2}}\, , \nonumber\\
J_{a_0}(x)&=&\frac{D^a(x)\bar{D}^a(x)-U^a(x)\bar{U}^a(x)}{\sqrt{2}}\, , \nonumber \\
J_{\kappa}(x)&=& S^a(x)\bar{U}^a(x) \, , \nonumber \\
J_{\sigma}(x)&=&S^a(x)\bar{S}^a(x) \, ,
\end{eqnarray}
where
\begin{eqnarray}
U^a(x)&\propto&\widetilde{U}^a(x)=\epsilon^{abc} d_b^T(x)C\gamma_5 s_c(x)\, , \nonumber\\
D^a(x)&\propto&\widetilde{D}^a(x)=\epsilon^{abc} u_b^T(x)C\gamma_5 s_c(x)\, ,\nonumber\\
S^a(x) &\propto& \widetilde{S}^a(x)=\epsilon^{abc} u_b^T(x)C\gamma_5
d_c(x)\, ,
\end{eqnarray}
the $a,~b,~c$ are color indices, the $C$ is the charge conjugation
matrix, the $U^a(x)$, $D^a(x)$ and $S^a(x)$ are basic scalar diquark
fields, while the $\widetilde{U}^a(x)$, $\widetilde{D}^a(x)$ and
$\widetilde{S}^a(x)$ are the corresponding  scalar two-quark
currents. In this article, we take the isospin limit for the $u$ and
$d$ quarks, and denote the fields $U^a(x)$ and $D^a(x)$ as $Q^a(x)$.

For the general  color antitriplet  bilinear quark-quark fields
$q(x)q(y)$\footnote{If we take the local limit in the scalar
channels, the two-quark currents $\widetilde{U}^a(x)$,
$\widetilde{D}^a(x)$ and $\widetilde{S}^a(x)$ are recovered.} and
color singlet bilinear
 quark-antiquark fields $q(x)\bar{q}(y)$, where the flavor, color and
spin indexes are not shown explicitly for simplicity, we can project
them into a local and a nonlocal part, after bosonization, the two
parts are translated into a basic quantum field and a bound state
amplitude, respectively,
\begin{eqnarray}
q(x)q(y)&\to&\mathbb{D}\left(\frac{x+y}{2}\right)\Gamma_{\mathbb{D}}(x-y)\, , \nonumber \\
 q(x)\bar{q}(y)&\to&\mathbb{M}\left(\frac{x+y}{2}\right)\Gamma_{\mathbb{M}}(x-y)\, ,
\end{eqnarray}
where the $\mathbb{D}(x)$ and $\mathbb{M}(x)$ denote the diquark and
meson fields respectively, the $\Gamma_{\mathbb{D}}(x)$ and
$\Gamma_{\mathbb{M}}(x)$ denote the corresponding Bethe-Salpeter
amplitudes respectively \cite{Tandy1997,Cahill1989}. In
Ref.\cite{WangCTP}, we study the structures of the pseudoscalar
mesons $\pi$, $K$ and the scalar diquarks $U^a$, $D^a$, $S^a$ in the
framework of the coupled rainbow Schwinger-Dyson equation and ladder
Bethe-Salpeter equation using a confining effective potential, and
observe that the dominant Dirac spinor  structure of the
Bethe-Salpeter amplitudes of the scalar diquarks  is $C\gamma_5$. If
we take the local limit for the nonlocal Bethe-Salpeter amplitudes,
the dimension-1 scalar diquark fields $U^a$, $D^a$ and $S^a$ are
proportional to the dimension-3 scalar two-quark currents
$\widetilde{U}^a$, $\widetilde{D}^a$ and $\widetilde{S}^a$,
respectively. A dimension-1 quantity $\Lambda$ can be introduced to
represent the hadronization $\widetilde{U}^a\approx\Lambda^2U^a$,
$\widetilde{D}^a\approx\Lambda^2D^a$ and
$\widetilde{S}^a\approx\Lambda^2S^a$.

 The attractive interaction
of one-gluon exchange favors  formation of the diquarks in  color
antitriplet $\overline{{\bf 3}}_{ c}$, flavor antitriplet
$\overline{{\bf 3}}_{ f}$ and spin singlet ${\bf 1}_s$
\cite{ReviewScalar,GI1}. Lattice QCD studies of  the light flavors
indicate that the strong attraction in the scalar diquark channels
favors the formation of  good diquarks, the weaker attraction (the
quark-quark correlation is rather weak) in the axial-vector diquark
channels maybe form bad diquarks, the energy gap between the
axial-vector and scalar diquarks is about $\frac{2}{3}$ of the
$\Delta$-nucleon mass splitting, i.e. $\approx 0.2\,\rm{GeV}$
\cite{Latt}, which is also expected from the hypersplitting
color-spin interaction $\frac{1}{m_im_j}\vec{T}_{i}\cdot \vec{T}_{j}
\vec{\sigma}_i \cdot \vec{\sigma}_j$ \cite{ReviewScalar,GI1}.  On
the other hand, the studies based on the random instanton liquid
model indicate that the instanton induced quark-quark interactions
are weakly repulsive in the vector and axial-vector channels,
strongly repulsive in the pseudoscalar channel, and strongly
attractive in the scalar and tensor channels \cite{RILM1994}. So it
is sensible to use the scalar diquark fields to construct the
tetraquark currents.

The  two-point correlation functions $\Pi_i(p)$ can be written as
\begin{eqnarray}
\Pi_i(p)=i\int d^4x ~e^{ip.x}\langle 0
|T[J_i(x){J_i}^\dagger(0)]|0\rangle \, ,
\end{eqnarray}
where the  current  $J_i(x)$ denotes $J_{f_0}(x)$,  $J_{a_0}(x)$,
  $J_{\kappa}(x)$ and $J_{\sigma}(x)$.

We can insert  a complete set of intermediate hadronic states with
the same quantum numbers as the current operators $J_i(x)$  into the
correlation functions  $\Pi_{i}(p)$  to obtain the hadronic
representation \cite{Shifman79,Reinders85}. Isolating the ground
state contributions from the pole terms of the nonet scalar mesons,
we obtain  the results,
\begin{eqnarray}
\Pi_i(p)&=&\frac{\lambda_i^{2}}{M_i^{2}-p^2}+\sum_h \frac{\langle 0|J_i(0)|h(p)\rangle \langle h(p)|J^{\dagger}_i(0)|0\rangle}{M_h^2-p^2} \, , \nonumber \\
&=&\lambda_i^{2} \int_{\Theta_i^2}^{s_i^0} ds
\frac{\delta\left(s-M_i^2\right)}{s-p^2}+\int_{s^0_i}^{\infty} ds
\frac{\rho^h_i(s)}{s-p^2}\, ,
\end{eqnarray}
where the $M_i$ are the ground state masses, the $\lambda_i$ are
corresponding pole residues defined by  $ \langle 0 |
J_i(0)|S(p)\rangle =\lambda_i$,  the thresholds
$\Theta_i^2=(2m_s+2m_q)^2, \, (m_s+3m_q)^2, \,(4m_q)^2$ in  the
channels $f_0/a_0(980)$, $\kappa(800)$, $\sigma(600)$ respectively,
the $s^0_i$ are the thresholds for the higher resonances and
continuum states $|h\rangle$,  and the $\rho^h_i(s)$ are the
corresponding hadronic spectral densities.

We introduce the following new Lagrangian $\mathcal {L}$,
\begin{eqnarray}
\mathcal {L}&=&\frac{1}{2}{\mathcal{D}_\mu U
}^\dagger\mathcal{D}^\mu U+\frac{1}{2}{\mathcal{D}_\mu D
}^\dagger\mathcal{D}^\mu D+\frac{1}{2}{\mathcal{D}_\mu S
}^\dagger\mathcal{D}^\mu S-\frac{1}{2}m_U^2{ U }^\dagger
U  -\frac{1}{2}m_D^2{ D }^\dagger D \nonumber \\
&& -\frac{1}{2}m_S^2{ S }^\dagger S+\cdots \, ,
\end{eqnarray}
where $\mathcal{D}_\mu=\partial_\mu-ig_sG_\mu$, and carry out the
operator product expansion with the basic diquark fields $U$, $D$
and $S$ instead of the quark fields $u$, $d$ and $s$, and we have
neglected the terms concerning  the heavy diquark fields in the
 Lagrangian. In the QCD, the basic quantum fields are the quark and
gluon fields, the attractive interactions in  the color antitriplet
$\overline{{\bf 3}}_{ c}$, flavor antitriplet $\overline{{\bf3}}_{
f}$ and spin singlet ${\bf 1}_s$ quark-quark channels favor
formation of the scalar diquarks, we can absorb some effects of the
quark-gluon interactions into the effective diquark masses, which
are characterized  by the correlation length $\mathbb{L}\sim
\frac{1}{\mathbb{D}}$. At the distance $l>\mathbb{L}$, the
$\bar{\bf{3}}_c$ diquark state combines with one quark or one
$\bf{3}_c$ antidiquark to form a baryon state or a tetraquark state,
while at the distance $l<\mathbb{L}$, the
 $\bar{\bf{3}}_c$ diquark states dissociate into asymptotic quarks and gluons gradually, the strength of the
   quark-quark correlations is very weak.
 Just like the quarks, the diquarks have
three colors, and can be gauged with the same $SU(3)$ color group to
embody the residual quark-gluon interactions.

\begin{figure}
 \centering
 \includegraphics[totalheight=8cm,width=14cm]{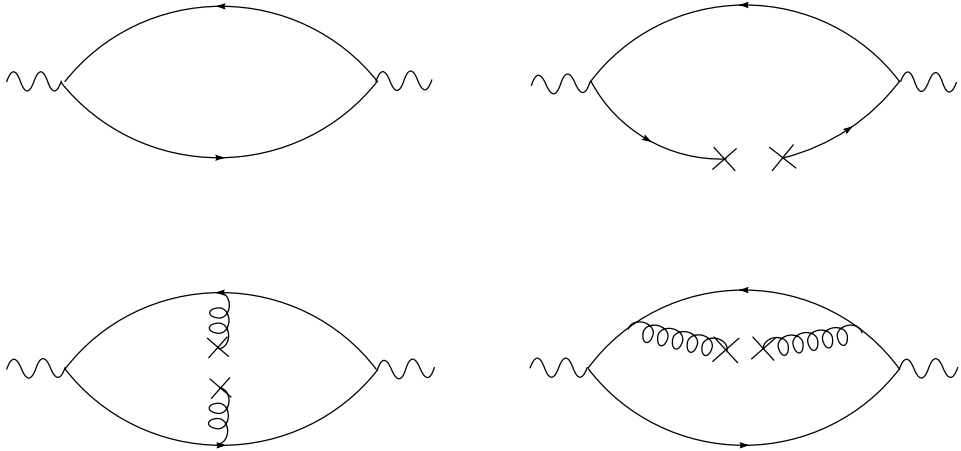}
    \caption{The typical diagrams we calculate in the operator product expansion,
    where the solid line denotes the diquark propagators. }
\end{figure}

In calculations, we take into account all diagrams like the typical
ones shown in Fig.1, introduce new vacuum diquark condensates
$\langle \bar{Q}Q\rangle$ and $\langle \bar{S}S\rangle$ besides the
gluon condensate to parameterize the nonperturbative QCD vacuum, and
consider the vacuum condensates up to dimension four. In the QCD sum
rules, the high dimensional vacuum condensates are usually
suppressed by  large denominators or additional powers of the
inverse Borel parameters $\frac{1}{T^2}$, the net contributions are
very small. For example, in the present case the contributions of
the dimension-6 vacuum condensates can be estimated as
$\langle\bar{Q}Q\rangle\langle \frac{\alpha_s
GG}{\pi}\rangle\frac{1}{T^{6}}\approx
 \frac{0.000037\,\rm{GeV}^6}{T^6}$, which is a tiny  quantity. If
 additional suppressions originate from the denominators are taken into account,
 the contributions are even smaller, and can be safely neglected.

  Once  the analytical  results are obtained,
  then we can take the dualities below the thresholds
$s_0$ and perform the Borel transform  with respect to the variable
$P^2=-p^2$, finally we obtain  the following sum rules,
\begin{eqnarray}
\lambda_i^{2}e^{-\frac{M_{i}^{2}}{T^2}}&=&\int_{\Delta_i^2}^{s_i^0}ds
e^{-\frac{s}{T^2}}\rho_i(s) \, ,
\end{eqnarray}
where the $\rho_i(s)$ denote  the QCD spectral densities
$\rho_{f_0/a_0}(s)$, $\rho_{\sigma}(s)$ and $\rho_{\kappa}(s)$,
\begin{eqnarray}
\rho_{f_0/a_0}(s)&=&\frac{3}{16\pi^2}+2\langle
\bar{Q}Q\rangle\delta(s-m_Q^2)-\frac{m_Q^2}{96T^4}\langle\frac{\alpha_sGG}{\pi}\rangle\int_0^1dx\frac{1}{x^3}
\delta(s-\widetilde{m}_Q^2) \, , \nonumber \\
\rho_{\sigma}(s)&=&\frac{3}{16\pi^2}+2\langle
\bar{S}S\rangle\delta(s-m_S^2)-\frac{m_S^2}{96T^4}\langle\frac{\alpha_sGG}{\pi}\rangle\int_0^1dx\frac{1}{x^3}
\delta(s-\widetilde{m}_S^2) \, ,\nonumber \\
\rho_{\kappa}(s)&=&\frac{3}{16\pi^2}+\langle
\bar{Q}Q\rangle\delta(s-m_S^2)+\langle
\bar{S}S\rangle\delta(s-m_Q^2)   \nonumber \\
&&-\frac{1}{192T^4}\langle\frac{\alpha_sGG}{\pi}\rangle\int_0^1dx
 \left[\frac{m_Q^2}{x^3}+\frac{m_S^2}{(1-x)^3}\right]\delta(s-\widehat{m}_{Q/S}^2)\, ,
\end{eqnarray}
 $\widetilde{m}_{S/Q}^2=\frac{m_{S/Q}^2}{x(1-x)}$, $\widehat{m}_{S/Q}^2=\frac{m_Q^2}{x}+\frac{m_S^2}{1-x}$,
  $\Delta_i^2=(2m_Q)^2,\,(m_Q+m_S)^2, \, (2m_S)^2$ in  the channels $f_0/a_0(980)$,  $\kappa(800)$, $\sigma(600)$
  respectively,  and the $T^2$ is the Borel
  parameter. The threshold parameters $\Delta_i^2$ are different
  from the corresponding  $\Theta_i^2$ in Eq.(6), because we absorb
  some QCD interactions into the effective diquark masses.

   Differentiate  Eq.(8) with respect to  $\frac{1}{T^2}$, then eliminate the
 pole residues $\lambda_{i}$, we can obtain the sum rules for
 the masses  of the nonet scalar mesons,
 \begin{eqnarray}
 M_{i}^2=\frac{ \int_{\Delta_{i}^2}^{s^0_i} ds \frac{d}{d(-1/T^2)}
\rho_i(s)e^{-\frac{s}{T^2}} }{\int_{\Delta_{i}^2}^{s^0_i} ds
\rho_i(s)e^{-\frac{s}{T^2}} }\, .
\end{eqnarray}

\section{Numerical Results}

We estimate the vacuum diquark condensates $\langle \bar{Q}Q\rangle$
and $\langle \bar{S}S\rangle$ with the assumption of the vacuum
saturation, which works well in the large $N_c$ limit,
\begin{eqnarray}
 \langle\overline{\widetilde{Q}} \widetilde{Q}\rangle&=&\frac{\langle\bar{q}q\rangle\langle\bar{s}s\rangle}{6}=\Lambda^4\langle\bar{Q}Q\rangle
\, ,  \nonumber \\
 \langle\overline{\widetilde{S}} \widetilde{S}\rangle&=&\frac{\langle\bar{q}q\rangle^2}{6}=\Lambda^4\langle\bar{S}S\rangle
 \, ,
\end{eqnarray}
where the $\Lambda$ is a quantity has dimension of mass and can be
taken as the confinement energy scale $\Lambda=0.3\,\rm{GeV}$. At
the energy scale $\mu=1\,\rm{GeV}$, $\langle \bar{s}s
\rangle=0.8\langle \bar{q}q \rangle$,  $\langle \bar{q}q
\rangle=-(0.24 \,\rm{GeV})^3$,  $\langle \frac{\alpha_sGG}{\pi}
\rangle=(0.33 \,\rm{GeV})^4$ \cite{Shifman79,Reinders85}, and
$\langle\bar{Q}Q\rangle=0.8\langle\bar{S}S\rangle=0.0031\,\rm{GeV}^2$.

The quark condensates play a special role being responsible for the
spontaneous breaking of the chiral symmetry, and relate with the
masses and decay constants of the light pseudoscalar mesons through
the Gell-Mann-Oakes-Renner relation \cite{GMOR}. The values of other
vacuum condensates, such  as the mixed condensates, the four quark
condensates and the gluon condensates, cannot be obtained from the
first principles,  we usually calculate them with the lattice QCD,
the instanton models, or determine them empirically by fitting the
QCD sum rules to the experimental data. In this article, we
introduce the diquark condensates $\langle \bar{Q}Q\rangle$ and
$\langle \bar{S}S\rangle$ to parameterize  the nonperturbative QCD
vacuum, and assume that they relate with the four quark condensates
(therefore they have implicit relations with the spontaneous
breaking of the chiral symmetry), and take the dimension-one
parameter $\Lambda$ to be the confinement energy scale, as the
 scalar mesons  $f_0(980)$, $a_0(980)$, $\kappa(800)$ and
$\sigma(600)$ are bound states which consist of the confined quarks
and gluons, and some parameters are needed to embody the
confinement. On the other hand, we can understand the parameter
$\Lambda=0.3\,\rm{GeV}$ as a fitted value, which happens to be the
confinement energy scale, the crude estimation works well.

 We take the updated values of the diquark masses  from the
QCD sum rules for consistency,
 where the interpolating currents $\widetilde{U}^a(x)$, $\widetilde{D}^a(x)$
and $\widetilde{S}^a(x)$ are used \cite{HuangDiquark},
$m_Q=0.46\,\rm{GeV}$ and $m_S=0.40\,\rm{GeV}$; the scalar diquarks
were originally studied with the QCD sum rules about twenty  years
ago \cite{diquark-SR}. There have been several theoretical
approaches to estimate the diquark masses, for example, the simple
constituent diquark mass plus hyperfine spin-spin interaction model
\cite{Maiani04}.

The $f_0(980)$ and $a_0(980)$ are well established, and the
existence of the $\sigma(600)$ meson is
  confirmed, although there are  controversy  about its mass and width, the
  values listed in the  Review of Particle Physics are
  $(400-1200)\,\rm{MeV}$ and $(600-1000)\,\rm{MeV}$ respectively \cite{PDG}.
  As far as the $\kappa(800)$  are concerned, there are still
  controversy about their existence, we take them as the
 $S$-wave isospin-$\frac{1}{2}$
  $K\pi$ resonance with the Breit-Wigner mass about $850\,\rm{MeV}$.
The E791 collaboration observed a low-mass scalar $K\pi$ resonance
with the Breit-Wigner mass $M=(797\pm 19 \pm 43) \,\rm{MeV}$ and
width $\Gamma=(410\pm 43\pm87) \,\rm{MeV}$ respectively in the decay
$D^+ \to K^- \pi^+ \pi^+$ \cite{E791}, and the BES collaboration
observed a clear low mass enhancement in the invariant $K\pi$ mass
distribution in the decay $J/\psi \to \bar{K}^*(892)K^+\pi^-$ with
the Breit-Wigner mass $M=(878 \pm 23^{+64}_{-55}) \,\rm{MeV}$ and
width $\Gamma=(499 \pm 52^{+55}_{-87})\,\rm{ MeV}$, respectively
\cite{BES-2005}. Recently, the BES collaboration reported the
charged $\kappa(800)$ in the decay  $J/\psi \to K^*(892)^{\mp} K_s
\pi^\pm$ with the Breit-Wigner mass $M=(826\pm49_{-34}^{+49})
\,\rm{MeV}$ and width $\Gamma=(449\pm156_{-81}^{+144})\,\rm{MeV}$,
respectively \cite{BES-1008}. It is sensible to
  estimate    $M_{f_0/a_0}-M_{\kappa}=m_s-m_q=0.14\,\rm{GeV}$. On
   the other hand, the QCD sum rules for the tetraquark states
   indicate that $M_{\kappa}=(0.80-0.88)\,\rm{GeV}$ and
   $M_{\sigma}=(0.72-0.80)\,\rm{GeV}$ \cite{WangYW}.

   Assuming the energy gap between the ground and first
   radial excited tetraquark states is about $0.5\,\rm{GeV}$,
   we can tentatively determine  the
    threshold parameters
$s^0_{f_0/a_0}=(1.0+0.5)^2\,\rm{GeV}^2$,
$s^0_{\kappa}=(0.85+0.5)^2\,\rm{GeV}^2$, and
 $s^0_{\sigma}=(0.75+0.5)^2\,\rm{GeV}^2$.

The convergence behavior of the operator product expansion is very
good, the contributions from the different terms have the hierarchy:
perturbative-term $>\langle\bar{Q}Q\rangle \gg \langle
\frac{\alpha_sGG}{\pi}\rangle$. In calculation, we take  uniform
minimum value for the Borel parameters
$T^2_{min}=\mu^2=1.0\,\rm{GeV}^2$. The perturbative continuum
$\frac{3}{16\pi^2}\Theta(s-s_0)$ is suppressed by the factor
$e^{-\frac{s}{T^2}}$,  the contributions from the pole terms are
very large, see Fig.2. In this article, we take uniform  maximum
value for the Borel parameters, $T^2_{max}=1.9\,\rm{GeV}^2$,  the
contributions from the pole terms are about $(61-84)\%$, $(54-79)\%$
and $(51-75)\%$ in the channels $f_0(980)/a_0(980)$, $\kappa(800)$
and $\sigma(600)$, respectively. The two criteria (pole dominance
and convergence of the operator product expansion) of the QCD sum
rules are well satisfied.

If the conventional quark currents are chosen, the multiquark states
i.e. tetraquark states, pentaquark states, hexaquark states, etc,
have the spectral densities $\rho\sim s^n$ with the largest
$n\geq4$, the integral $\int_0^{\infty} s^n e^{-\frac{s}{T^2}} ds$
converges slowly \cite{Narison04}. If one don't want to release the
criterion of pole dominance, we have to either postpone the
threshold parameter $s_0$ to very large value or choose very small
value for the Borel parameter $T^2_{max}$. With large value of the
threshold parameter $s_0$ , for example, $s_0 \gg M_{\rm gr}^2$,
here $\rm{gr}$ stands for the ground state, the contributions from
the high resonance states and continuum states are included in, we
cannot use single-pole (or ground state) approximation for the
spectral densities; on the other hand, with very small value of the
Borel parameter $T^2_{max}$,  the Borel window $T^2_{max}-T^2_{min}$
shrinks to zero or very small values.

The numerical values of the masses and pole residues are presented
in Table 1 and Figs.3-4. From Table 1, we can see that the present
predictions are compatible with (or not in conflict with) the
experimental data \cite{PDG,E791,BES-2005,BES-1008} and theoretical
estimations \cite{WangYW}.

The scalar tetraquark currents $J_{f_0/a_0}(x)$, $J_{\kappa}(x)$ and
$J_{\sigma}(x)$ maybe have non-vanishing couplings  with the
scattering states $\pi\pi$, $K\bar{K}$, $K\pi$, $K\eta$, $\pi\eta$,
$\eta\eta$, etc, for example,
\begin{eqnarray}
\langle 0|J_{f_0}(0)|\pi\pi(p)\rangle &=&\lambda_{f_0\pi\pi} \, , \nonumber \\
 \langle 0|J_{f_0}(0)|K\bar{K}(p)\rangle &=&\lambda_{f_0KK} \, .
\end{eqnarray}
If the couplings  denoted by the $\lambda_{f_0\pi\pi}$ and
$\lambda_{f_0KK}$ are strong enough, the contaminations from the
continuum states are expected to be large. In the following, we
study the contributions of the  intermediate pseudoscalar meson
loops to the correlation function $\Pi_{f_0}(p)$ in details as an
example,
\begin{eqnarray}
\Pi_{f_0}(p)&=&\frac{\lambda_{f_0}^{2}}{M_{f_0}^{2}-p^2}-i\frac{\lambda_{f_0}}{p^2-M_{f_0}^2}
g_{f_0\pi\pi}\Sigma_{\pi\pi}(p)g_{f_0\pi\pi}\frac{\lambda_{f_0}}{p^2-M_{f_0}^2}  \nonumber \\
&&-i\frac{\lambda_{f_0}}{p^2-M_{f_0}^2}g_{f_0KK}\Sigma_{KK}(p)g_{f_0KK} \frac{\lambda_{f_0}}{p^2-M_{f_0}^2}\nonumber \\
&&-\frac{\lambda_{f_0}}{p^2-M_{f_0}^2}g_{f_0\pi\pi}\Sigma_{\pi\pi}(p)T_{\pi\pi
\to\pi\pi}\Sigma_{\pi\pi}(p)g_{f_0\pi\pi}\frac{\lambda_{f_0}}{p^2-M_{f_0}^2}\nonumber \\
 &&-\frac{\lambda_{f_0}}{p^2-M_{f_0}^2}g_{f_0\pi\pi}\Sigma_{\pi\pi}(p)T_{\pi\pi \to K \bar{K}}\Sigma_{KK}(p)g_{f_0KK}\frac{\lambda_{f_0}}{p^2-M_{f_0}^2}\nonumber \\
 &&-\frac{\lambda_{f_0}}{p^2-M_{f_0}^2}g_{f_0KK}\Sigma_{KK}(p)T_{K\bar{K}\to\pi\pi}\Sigma_{\pi\pi}(p)g_{f_0\pi\pi}\frac{\lambda_{f_0}}{p^2-M_{f_0}^2}\nonumber \\
 &&-\frac{\lambda_{f_0}}{p^2-M_{f_0}^2}g_{f_0KK}\Sigma_{KK}(p)T_{K\bar{K}\to K\bar{K}}\Sigma_{KK}(p)g_{f_0KK}\frac{\lambda_{f_0}}{p^2-M_{f_0}^2}+\cdots
\, ,
\end{eqnarray}
where
\begin{eqnarray}
\Sigma_{\pi\pi}(p)&=&\int~{d^4q\over(2\pi)^4}\frac{1}{\left[
q^2-m_{\pi}^2\right]\left[ (p-q)^2-m_{\pi}^2\right]} \, , \nonumber\\
\Sigma_{KK}(p)&=&\int~{d^4q\over(2\pi)^4}\frac{1}{\left[
q^2-m_{K}^2\right]\left[ (p-q)^2-m_{K}^2\right]} \, ,
\end{eqnarray}
the $g_{f_0\pi\pi}$, $g_{f_0KK}$ are the strong coupling constants
between  the $f_0(980)$ and the pseudoscalar meson pairs $\pi\pi$,
$K\bar{K}$ respectively, and the $T_{\pi\pi \to \pi\pi}$, $T_{\pi\pi
\to K\bar{K}}$, $T_{ K\bar{K} \to \pi\pi}$, $T_{ K\bar{K} \to
K\bar{K}}$ are the scattering amplitudes among  the pseudoscalar
meson pairs $\pi\pi$ and $K\bar{K}$. The couplings $\lambda_{f_0
\pi\pi}$ and $\lambda_{f_0KK}$ are complicated  functions of the
$M_{f_0}$, $p^2$, $\lambda_{f_0}$, $T_{\pi\pi \to \pi\pi}$,
$T_{\pi\pi \to K\bar{K}}$, $T_{ K\bar{K} \to \pi\pi}$, $T_{ K\bar{K}
\to K\bar{K}}$, $g_{f_0\pi\pi}$ and $g_{f_0KK}$, the explicit
expressions are difficult to obtain. We should bear in mind that the
intermediate meson loops contribute a self-energy to the scalar
meson $f_0(980)$, and therefore  the scalar meson $f_0(980)$
develops a Breit-Wigner width. In fact, the scalar mesons
$f_0/a_0(980)$, $\kappa(800)$ and $\sigma(600)$ below $1\,\rm{GeV}$
can be  generated dynamically from the unitaried scattering
amplitudes of the pseudoscalar mesons \cite{OsetReview}. We can take
into account those meson loops effectively  by taking the following
replacement for the hadronic spectral density,
\begin{eqnarray}
\delta\left(s-M_{f_0}^2\right) &\to& \frac{1}{\pi}
\frac{\sqrt{s}\Gamma_{f_0}}{\left(s-M_{f_0}^2
\right)^2+s\Gamma_{f_0}^2} \, ,
\end{eqnarray}
here we neglect the complicated renormalization procedure, take the
physical values, and ignore  the energy scale dependence of the mass
$M_{f_0}$ and pole residue $\lambda_{f_0}$ for simplicity; the
approximation works well. The QCD sum rules for other scalar mesons
are treated with the same routine. In Ref.\cite{ZhuEPJC}, Dai et al
perform the renormalization procedure in details to take into
account the contributions from the continuum states.

The widths listed in the Review of Particle Physics are
$\Gamma_{f_0}=(40-100)\,\rm{MeV}$,
$\Gamma_{a_0}=(50-100)\,\rm{MeV}$,
  $\Gamma_{\kappa}=(550\pm34)\,\rm{MeV}$ and $\Gamma_{\sigma}=(400-1200)\,\rm{MeV}$, respectively
  \cite{PDG}. Taking into account the finite widths, we can obtain the
  modified masses from the corresponding QCD sum rules, which are shown in Table 1
  and Fig.5. In calculation, we observe that the narrow width
  $\Gamma_{f_0/a_0}$ modifies   the mass $M_{f_0/a_0}$ slightly and
  the effect can be neglected safely, while the broad widths $\Gamma_{\kappa}$ and $\Gamma_{\sigma}$
  reduce the masses $M_{\kappa}$ and $M_{\sigma}$ about
  $55\,\rm{MeV}$ and $(0-75)\,\rm{MeV}$, respectively. Comparing with the experimental data
  \cite{PDG,E791,BES-2005,BES-1008}, the modified masses are better.

  The scalar tetraquark currents $J_{f_0/a_0}(x)$, $J_{\kappa}(x)$ and
$J_{\sigma}(x)$ maybe also have non-vanishing couplings with the
higher resonances $a_0(1450)$, $f_0(1370)$, $K^*(1430)$, $f_0(1500)$
and $f_0(1710)$, which  are supposed to be the ${}^3P_0$ $q\bar{q}$
states or glueballs, the couplings should be very small.
Furthermore, the threshold parameters are $s_0=2.25\,\rm{GeV}^2$,
$1.82\,\rm{GeV}^2$ and $1.56\,\rm{GeV}^2$ in the channels
$f_0/a_0(980)$, $\kappa(800)$ and $\sigma(600)$,  respectively, the
contaminations should be very small.

 \begin{table}
\begin{center}
\begin{tabular}{|c|c|c|}\hline\hline
           tetraquark states                             & mass (GeV)      &  pole residue (GeV) \\ \hline
          $f_0(980)/a_0(980)$                            & $0.94-1.04$     & $0.16-0.17$ \\ \hline
           $\kappa(800)$                                 & $0.84-0.93$     & $0.15-0.16$ \\ \hline
                  $\sigma(600)$                          & $0.77-0.84$     & $\approx0.15$ \\ \hline
   $f_0(980)/a_0(980)$ $\left[(40-100)\,\rm{MeV}\right]$ & $0.94-1.04^*$   &  \\ \hline
         $\kappa(800)$      $\left[550\,\rm{MeV}\right]$ & $0.79-0.88^*$   &   \\ \hline
           $\sigma(600)$    $\left[400\,\rm{MeV}\right]$ & $0.77-0.84^*$   &  \\ \hline
       $\sigma(600)$        $\left[800\,\rm{MeV}\right]$ & $0.72-0.80^*$   &  \\ \hline
         $\sigma(600)$     $\left[1200\,\rm{MeV}\right]$ & $0.69-0.77^*$   &  \\ \hline
     \hline
\end{tabular}
\end{center}
\caption{ The masses and the pole residues  of the  nonet $0^{++}$
tetraquark states. The star denotes the modified masses from the sum
rules where the finite widths shown in the bracket are taken into
account in the hadronic spectral densities. }
\end{table}

\begin{figure}
 \centering
 \includegraphics[totalheight=6cm,width=8cm]{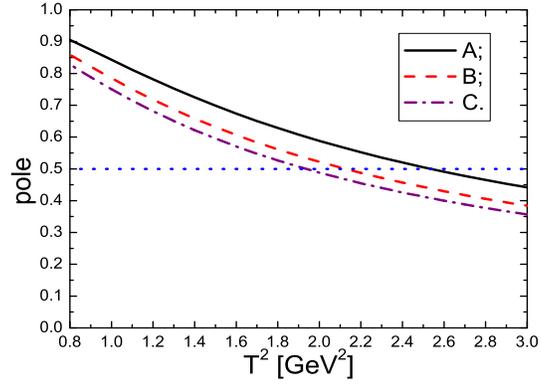}
    \caption{The contributions from the pole terms. The $A$,
   $B$ and $C$ correspond to the channels $f_0(980)/a_0(980)$,
$\kappa(800)$ and $\sigma(600)$ respectively. }
\end{figure}

\begin{figure}
 \centering
 \includegraphics[totalheight=6cm,width=8cm]{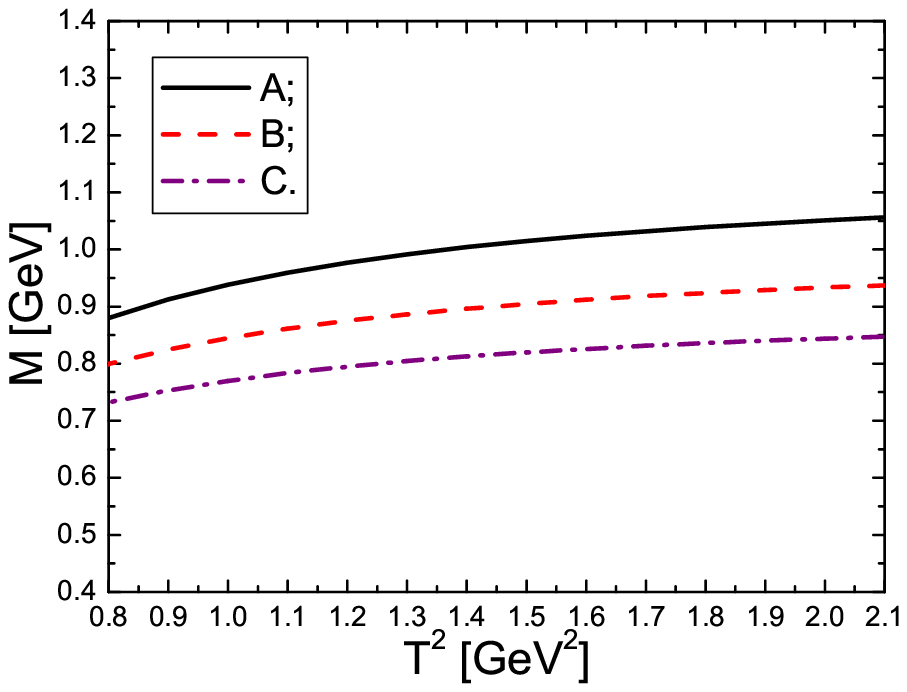}
    \caption{The masses   of the  nonet
    scalar tetraquark states. The $A$,
   $B$ and $C$ correspond to the channels $f_0(980)/a_0(980)$,
$\kappa(800)$ and $\sigma(600)$ respectively. }
\end{figure}

\begin{figure}
 \centering
 \includegraphics[totalheight=6cm,width=8cm]{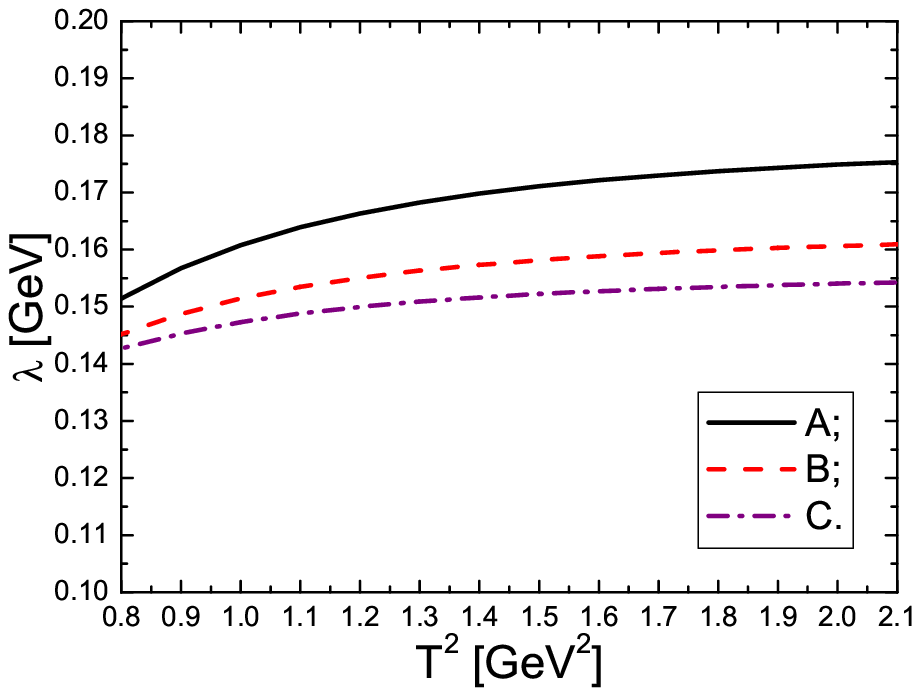}
    \caption{The  pole residues  of the  nonet
    scalar tetraquark states. The $A$,
   $B$ and $C$ correspond to the channels $f_0(980)/a_0(980)$,
$\kappa(800)$ and $\sigma(600)$ respectively. }
\end{figure}

\begin{figure}
 \centering
 \includegraphics[totalheight=6cm,width=8cm]{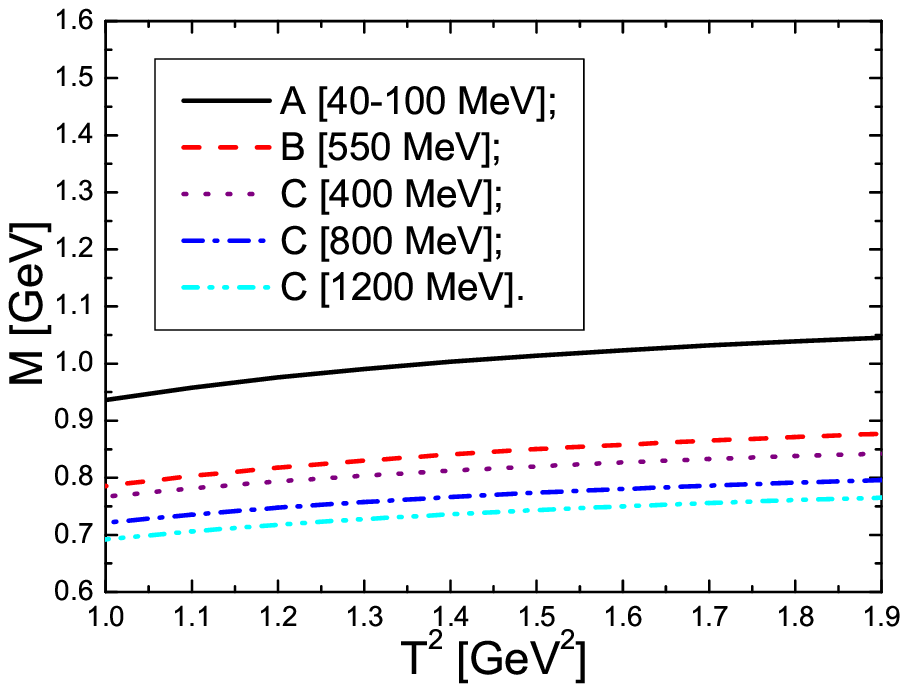}
    \caption{The modified  masses   of the  nonet
    scalar tetraquark states. The $A$,
   $B$  and $C$ correspond to the channels $f_0(980)/a_0(980)$,
$\kappa(800)$ and $\sigma(600)$ respectively. The values in the
bracket denote the finite widths in the hadronic spectral densities.
}
\end{figure}

\section{Conclusions}
In this article, we take the scalar diquarks as the point particles
and describe them as the basic quantum fields, then introduce the
$SU(3)$ color gauge interactions, and construct the tetraquark
currents which consist of the scalar fields to study the nonet
scalar mesons as tetraquark states with the new QCD sum rules. The
numerical values are compatible with (or not in conflict with) the
experimental data and theoretical estimations. Comparing with the
conventional quark currents, the diquark currents have the
outstanding advantage to
 satisfy the two criteria of the QCD sum rules more easily, the new
sum rules can be extended to study other multiquark states.

\section*{Acknowledgment}
This  work is supported by National Natural Science Foundation of
China, Grant Number 11075053,  and the Fundamental Research Funds
for the Central Universities.

\end{document}